\documentclass[aip,pof,12pt,preprint]{revtex4-1}
\usepackage{amssymb,amsbsy,times,fancyhdr,color}
\usepackage{amsmath,bm}
\usepackage{mathrsfs,amsfonts}
\usepackage{natbib}
\usepackage{latexsym,afterpage}
\usepackage{graphicx,subfigure,multirow}
\usepackage{color}

\newcommand{\imi}{\mathrm{i}}

\newcommand{\phiby}[1]{$\phi=\frac{\pi}{#1}$}

\begin{document}


\title{Mean flow generation in rotating anelastic two-dimensional convection} 

%

\author{Laura K. Currie}
\email[]{lcurrie@astro.ex.ac.uk}
\affiliation{Department of Physics and Astronomy, University of Exeter, Exeter, EX4 4QL,
UK}
\affiliation{Department of Applied Mathematics, University of Leeds, Leeds, LS2 9JT, UK}

\author{Steven M. Tobias}
\affiliation{Department of Applied Mathematics, University of Leeds, Leeds, LS2 9JT, UK}

%
\date{18 December 2015}

\begin{abstract}
We investigate the processes that lead to the generation of mean flows in two-dimensional anelastic convection. The simple model consists of a plane layer that is rotating about an axis inclined to gravity. The results are two-fold: firstly we numerically investigate the onset of convection in three-dimensions, paying particular attention to the role of stratification and highlight a curious symmetry. Secondly, we investigate the mechanisms that drive both zonal and meridional flows in two dimensions. We find that in general non-trivial Reynolds stresses can lead to systematic flows and, using statistical measures, we quantify the role of stratification in modifying the coherence of these flows.
\end{abstract}


\maketitle 



\section{Introduction}

Geophysical and astrophysical flows are often turbulent and characterised by the presence of a wide-range of temporal and spatial scales. It is often the case that systematic large-scale flows that vary on long timescales co-exist with shorter lived turbulent eddies. Famous examples of such large-scale flows are the zonal jets so visible at the surface of the gas giants \cite{Porco2003, Vasavada2005},  the turbulent jet stream of the Earth \cite{Vallis2006} and the strong zonal and meridional flows in the interior of the Sun which are interpreted as the observed differential rotation and meridional circulations\cite{Schouetal1998}.

A central question for fluid dynamicists is then to determine the role of the smaller-scale turbulent flows in modifying the systematic (or mean) flows. In some cases the turbulence will act purely as a dissipation or friction and act so as to damp any mean flows that occur. However in certain circumstances (usually when rotation is important) the turbulence may act as an anti-friction \cite{Mcintyre2003} and play a key role in driving and maintaining the mean flows. This mechanism is complicated by the tendency of the mean flows to act back on the turbulence and modify its form; a process that often determines the saturation amplitude of the mean flows. These are complicated interactions and although much progress has been made (as discussed briefly below) there is still much that is not understood.

Here we focus on a simple model where the turbulence is driven by a thermal gradient leading to convection. Owing to its importance in planetary and stellar interiors, this has been an extremely well studied problem, with many theoretical and numerical studies performed both in Cartesian (local) \cite{HS1983,HS1986,HS1987,JulienKnobloch,SaitoIshioka2011} and spherical (both shell and full sphere) geometries \cite{Miesch2000,Elliottetal2000,Christensen2001,Christensen2002,BrunToomre2002,Browningetal2004,GastineWicht2012,Gastineetal2013}. We note that, in order to capture some effects of vortex stretching in a spherical body, \citet{Busse1970} introduced an annulus model for its relative simplicity. This geometry has been used in attempts to model the zonal flow on Jupiter. For example, \citet{Jonesetal2003} used a rotating annulus model in a two-dimensional (2d) study and incorporated the possibility of boundary friction which allowed for the more realistic multiple jet solutions to be found more easily. \citet{RotvigJones2006} examined this annulus model more extensively and identified a bursting mechanism that occurs in the convection in some cases.

We wish to focus on the role of stratification in altering the dynamics of the mean flow and so focus on a simple plane layer model in two dimensions allowing us to access some parameter regimes more easily than in other, more complicated geometries. The plane layer model, when the axis of rotation is allowed to vary from the direction of gravity, can be used to represent a local region at different latitudes of a spherical body. This paper builds on previous, largely Boussinesq, studies in a Cartesian domain which are summarised here - though this is by no means a complete review.

Much of the previous work using a local model makes the Boussinesq approximation so that density variations are neglected except in the buoyancy force \cite{Boussinesq1903,SpiegelVeronis}. For example, \citet{JulienKnobloch} studied nonlinear convection cells in a rapidly rotating fluid layer with a tilted rotation vector. They performed an asymptotic analysis and concluded the orientation of the convection rolls affects the efficiency of mean flow generation. \citet{HS1983} performed three-dimensional (3d) simulations of convection in layers with tilted rotation vectors and no slip boundary conditions. They found that the horizontal component of the rotation vector gives a preference for cells aligned with the rotation axis which produces dynamical changes that drive a mean flow. There have also been a number of studies analysing the role of shear flow in a rotating plane layer. \citet{HS1987} considered Boussinesq convection in a rotating system with an imposed shear flow. The imposed flow was constant in depth, but varying in latitude. They found in the non-rotating case that the convection extracts energy from the mean flow and reduces the shear, but in the rotating case the convection can feed energy into the mean flow and increase the shear. \citet{HS1986} studied the interaction between convection, rotation and flows with vertical shear. They performed 3d simulations with no slip boundary conditions and found that in cases with vertical rotation the convection becomes more energetic by extracting energy from the mean flow. However, for cases with a tilted rotation vector the results depend on the direction of the shear. \citet{SaitoIshioka2011} revisited the problem of the interaction of convection with rotation in an imposed shear flow. They were able to examine a larger region of parameter space than \citet{HS1987} and identified a feedback mechanism in which the convection interacts with the rotation in such a way that leads to an accelerated mean flow. This mechanism operates when the sign of the shear flow is opposite to the vertical component of the rotation axis and relies upon the sinusoidal form of shear flow they imposed. \citet{mythesis} studied the generation of mean flows by Reynolds stresses in Boussinesq convection both in the absence and in the presence of a thermal wind and showed whether convection acts to increase or decrease the thermal wind shear depends on the fluid Prandtl number and the angle of the rotation vector from the vertical.

In many astrophysical systems the fluid is strongly stratified (e.g., in stellar and some planetary interiors) and therefore density changes across the layer (that are neglected by the Boussinesq approximation) may play a significant role. Therefore, there exist studies where fully compressible convection has been simulated \cite{Brummelletal1996,Brummelletal1998,Chan2001} but the full equations are computationally intensive to solve owing to the necessity of accurately tracking sound waves.
However, for systems where there are a large number of scale heights involved but that remain close to being adiabatic, the anelastic equations are an improvement on the Boussinesq equations \cite{Gough1969, LantzFan1999, BragRob1995}. Furthermore, the anelastic equations allow for density stratification across the layer whilst still filtering out fast sound waves thus making studying a compressible layer more computationally accessible.

In this paper we use the anelastic approximation to investigate the effects of stratification on both linear and nonlinear convection, focussing on the role stratification plays in altering the dynamics of mean flows. In other words, we extend the work of \citet{HS1983} to include the important effects of stratification.

By considering density variations across the fluid layer one cannot take both the dynamic viscosity $\mu$ and the kinematic viscosity $\nu$ of the fluid to be constant (since $\mu=\rho \nu$ where $\rho$ is the fluid density) therefore, a choice has to made. Similarly, one cannot take both the thermal conductivity $k$ and the thermal diffusivity $\kappa$ of the fluid to be constant and again, a choice has to be made. The results can depend on these choices (see, e.g., \citet{GlatzmaierGilman1981}). In this paper we consider a formalism where $\nu$ and $k$ are constant.
Anelastic formalisms also differ depending on whether entropy or temperature is diffused in the energy equation\cite{BragRob1995}. If one diffuses entropy then temperature can be eliminated as a variable from the formulation; for simplicity, this is often the approach taken in nonlinear studies and is the choice we make here. 
Further differences between anelastic formalisms arise from whether one takes the vertical axis to be parallel or antiparallel to gravity. Typically, in Boussinesq formulations the vertical axis increases upwards, whereas in compressible studies it is taken downwards. In order to ease the comparison with Boussinesq models, we take the vertical axis to increase upwards in line with \citet{Mizerski}. 
The existing studies of anelastic convection, linear and nonlinear, each consider a different formalism (discussed below) and therefore care has to be taken when comparing across the different models.

The onset of compressible convection in a local Cartesian geometry, using the anelastic approximation, has been studied in a number of papers. The earliest of these studies includes the work of \citet{KatoUnno1960} who studied the onset of convection in an isothermal reference atmosphere. As an alternative, it is common to assume a polytropic reference atmosphere, as we do here. For example, \citet{JonesRobertsGalloway1990} considered a polytropic, constant conductivity model in a Cartesian geometry in which rotation, magnetic field and gravity were taken to be mutually perpendicular. This is different to the configuration in this paper as we consider rotation that is oblique to gravity.
More recently, \citet{Mizerski} investigated the effect of compressibility and stratification on convection, using the anelastic approximation, in a rotating plane layer model where rotation and gravity were aligned; they compared a model with constant $k$ with one with constant $\kappa$ whilst keeping $\nu$ constant throughout and they chose to diffuse entropy. In this paper we essentially consider the constant conductivity case of \citet{Mizerski} but we allow the rotation vector to tilt from the vertical in order to model behaviour at mid-latitudes of a spherical body.
To date, the only other study examining the linear stability of compressible convection in this tilted f-plane geometry is that of \citet{Calkins2014} who compared the onset of convection in compressible and anelastic ideal gases on a tilted f-plane. The majority of the results from that paper are concerned with comparing the anelastic equations with temperature diffusion with the fully compressible equations, and so those results are not directly comparable with ours. They do however, also consider an entropy diffusion model but in contrast to the work here, they take $\mu$ constant and have the vertical axis pointing downwards.

In contrast to the Boussinesq systems, there have been relatively few studies of mean flow generation by convection under the anelastic approximation in a local model. 
\citet{RogersGlatzmaier2005} did model penetrative convection in a system where a convective region is bounded below by a stable region and \citet{Rogersetal2003} presented 2d simulations of turbulent convection using the anelastic approximation in a non-rotating system.
However, most relevant to the work we undertake here is the study of \citet{VerhoevenStellmach2014} who used 2d anelastic simulations of rapidly rotating convection in the equatorial plane (gravity and rotation perpendicular) to support a compressional Rhines-type mechanism predicting the width of jets driven in such a rapidly rotating system.
In their study, as we do,  \citet{VerhoevenStellmach2014} take $\nu$ constant and consider an entropy diffusion model, however, in contrast to our work,  \citet{VerhoevenStellmach2014} take $\kappa$ constant.

Our study, using a local model, allows us to focus on the effect of stratification on mean flow generation without many of the complicating features of a spherical geometry, say. However our model still incorporates some important physical features that are expected to play a role in the determining the dynamics of many astrophysical objects, namely stratification on the f-plane.

This paper is organised as follows: in section \ref{sec:model}, the model and governing equations are presented. In section \ref{sec:linear}, we aim to add to the existing literature discussed above by carrying out 3d linear stability analysis and present a new symmetry of which technical details are given in an appendix. In section \ref{sec:NL}, we analyse the role of stratification in mean flow generation at mid-latitudes using a 2d local model, focussing on the differences from Boussinesq models. This work complements previous Boussinesq and fully compressible studies in a Cartesian domain, as well as the global anelastic models considering mean flows.

\section{Model setup and equations}\label{sec:model}
We consider a local plane layer of convecting fluid rotating about an axis that is oblique to gravity, which acts downwards. The rotation vector lies in the $y$-$z$ plane and is given by $\bm{\Omega}=(0,\Omega\cos\phi,\Omega\sin\phi)$, where $\phi$ is the angle of the tilt of the rotation vector from the horizontal, so that the layer can be thought of as being tangent to a sphere at a latitude $\phi$. In this case, the $z$-axis points upwards, the $x$-axis eastwards and the $y$-axis northwards.



We denote the fluid density, pressure, temperature, entropy and velocity by $\rho$, $p$, $T$, $s$ and $\mathbf{u} = (u,v,w)$ respectively. The anelastic equations are then found by decomposing the thermodynamic variables into a reference state (denoted by the subscript $_{\rm{ref}}$) and a perturbation. The reference state depends on $z$ only and is assumed to be almost adiabatic. The departure from adiabaticity is
measured by a small parameter, $\epsilon$ given by
\begin{equation}\label{epsilondef}
 \epsilon\equiv\frac{d}{H_r}\left(\frac{\partial \ln  T_{\rm{ref}}}{\partial \ln 
p_{\rm{ref}}}-\frac{\partial \ln T_{\rm{ref}}}{\partial \ln 
p_{\rm{ref}}}\bigg|_{\text{ad}}\right)=-\frac{d}{T_r}\left[\left(\frac{d T_{\rm{ref}}}{dz}\right)_r+\frac{g}{c_p}\right]=-\frac{d}{c_p}\left(\frac{d s_{\rm{ref}}}{dz}\right)_r,
\end{equation}
where $d$ is the layer depth, $H_r=\frac{p_{\rm{ref}}}{g
\rho_{\rm{ref}}}=-\frac{{\rm d}z}{{\rm d}\ln p_{\rm{ref}}}$ is the pressure scale height, $c_p$ is the specific heat at constant pressure, $g$ is the acceleration due to gravity, the subscript
$_\text{ad}$ indicates the value for an adiabatic atmosphere and a subscript $_r$ denotes a value taken at the bottom of the layer, $z=0$.  $\epsilon$ will also be a
measure of the relative magnitude of the perturbations and we assume
\begin{equation}\label{pertratios}
 \frac{|p|}{p_{\rm{ref}}} \approx  \frac{|\rho|}{\rho_{\rm{ref}}} \approx  \frac{|T|}{T_{\rm{ref}}}
\approx |s| \approx \epsilon \ll 1,
\end{equation}
so that the perturbations are small compared to the reference state. 

It is useful to write the equations in a dimensionless form, using $d$ as the length scale and the thermal diffusion time $\frac{d^2}{\kappa_r}$ as the time scale, where $\kappa$ is the thermal diffusivity. In this case, the leading order equations are given by 
\begin{align}\label{newmom}
\left[\frac{\partial \mathbf{u}}{\partial t}+(\mathbf{
u}\cdot\nabla)\mathbf{ u} \right]=-\nabla\left(\frac{
p}{\rho_{\rm{ref}}}\right)+RaPrs\mathbf{\hat e_z}-Ta^{\frac{1}{2}}Pr\bm{
\Omega}\times\mathbf{
u}+\frac{Pr}{\rho_{\rm{ref}}}\nabla\cdot{\bm\varsigma},
\end{align}
\begin{equation}\label{newcont}
 \nabla\cdot(\rho_{\rm{ref}}{\mathbf{u}})=0,
\end{equation}
\begin{align}\label{newenergy}
\rho_{\rm{ref}} T_{\rm{ref}}\left[\frac{\partial{s}}{\partial  t}+(\mathbf{
u}\cdot\nabla)(s_{\rm{ref}}+ s)\right]
=\nabla\cdot[T_{\rm{ref}}\nabla(s_{\rm{ref}}+ s)]-\frac{\theta}{\rho_{\rm{ref}}
Ra}\frac{\bm{\varsigma}^{2}}{2},
\end{align}
\begin{equation}
 \frac{p}{p_{\rm{ref}}} = \frac{T}{T_{\rm{ref}}}+ \frac{\rho}{\rho_{\rm{ref}}},
\end{equation}
\begin{equation}\label{news}
s=\frac{1}{\gamma}\frac{p}{p_{\rm{ref}}} - \frac{\rho}{\rho_{\rm{ref}}},
\end{equation}
where $\varsigma_{ij}= \rho_{\rm{ref}} \left[ \frac{\partial u_i}{\partial x_j}+\frac{\partial u_j}{\partial x_i} - \frac{2}{3}(\nabla\cdot \mathbf u)\delta_{ij}\right]$ is the stress tensor with $\bm\varsigma^2\equiv \bm\varsigma:\bm\varsigma =\varsigma_{ij}\varsigma_{ij}$. $\theta$ is the dimensionless superadiabatic temperature gradient and $\gamma$ is the ratio of specific heats at constant pressure to constant volume. In this case the dimensionless parameters are the Rayleigh, Taylor and Prandtl numbers given by 
\begin{equation}
 Ra=\frac{gd^3}{\kappa_r\nu},
 \mbox{\quad }
 Ta=\frac{4\Omega^2d^4}{\nu^2}
\mbox{\quad and \quad}
     Pr=\frac{\nu}{\kappa_r},
    \end{equation}
respectively. Again, note in our formalism that we assume the kinematic viscosity $\nu$ and the turbulent thermal conductivity $k=\rho_{\rm{ref}} c_p\kappa$ (where we now interpret $\kappa$ as the turbulent thermal diffusivity) to be constant.
Furthermore, we use a model that takes the turbulent thermal conductivity to be much larger than
the molecular conductivity and so equation (\ref{newenergy}) contains an entropy diffusion term but not a thermal diffusion term; \citet{BragRob1995} discuss models including both terms. In addition, we have employed a technique introduced by \citet{Lantz1992} and \citet{BragRob1995} to reduce the number of thermodynamic variables\cite{LantzFan1999, JonesKuzanyan2009}.
%

In this paper, we consider a time-independent, polytropic reference state given by
\begin{align}\label{eq:basicstate}
T_{\rm{ref}}&=1+\theta z, \quad  \rho_{\rm{ref}}= (1+\theta z)^m, \quad 
p_{\rm{ref}}=-\frac{RaPr}{\theta(m+1)}(1+\theta z)^{m+1},\\
s_{\rm{ref}} &= \frac{m+1-\gamma m}{\gamma \epsilon}\ln(1+\theta z)+\text{const} \text{ \quad
with \quad} \frac{m+1-\gamma
m}{\gamma}=-\frac{\epsilon}{\theta}=\mathcal{O}(\epsilon),\label{eq:basicstateend}
\end{align}
where $m$ is the polytropic index and $-1 < \theta \le 0$.
We note in this model there is no adjustment of the reference state by any mean that may be generated.

The equations (\ref{newmom})-(\ref{news}) are similar to those given in case (1) of \citet{Mizerski}; the key difference here however is the introduction of a tilted rotation vector, $\mathbf{\Omega}=(0,\cos\phi, \sin\phi)$.
In this formalism, the anelastic equations (\ref{newmom})-(\ref{news}) reduce to the Boussinesq equations in the limit $\theta\rightarrow0$ and so $\theta$ can be thought of as a measure of the degree of compressibility.  However, a more intuitive measure is given by
\begin{align}
 \chi=\frac{\rho_{\rm{ref}}|_{\text{z=0}}}{ \rho_{\rm{ref}}|_{\text{z=1}}}=(1+\theta)^{-m}.
\end{align}
As $\chi$ is increased, the density contrast across the layer is also increased. Furthermore, it is common in the literature to use $N_{\rho}$, the number of density scale heights, as a measure of the stratification.
Note, $\theta$, $\chi$ and $N_{\rho}$ are related by the following relations:
$N_{\rho}=\ln \chi=\ln(1+\theta)^{-m}$.

With the reference state specified, we solve the anelastic equations subject to impenetrable, stress free and fixed entropy boundary conditions.




\section{Linear theory} \label{sec:linear}
Here we perform linear stability analysis of the 3d system in order to obtain information about the preferred modes of convection at onset.
We linearise the anelastic equations (\ref{newmom})-(\ref{newenergy}) by perturbing about a trivial basic state and neglecting terms quadratic in the perturbation quantities. We then assume the perturbations are given by
\begin{align}
 (s,w,\zeta)&={\rm Re}\left\{\left[S(z),W(z),Z(z)\right]e^{\imi(kx+ly)+\sigma t}\right\}.
\end{align}
Here $k$ and $l$ are the $x$ and $y$ wavenumbers respectively such that $a^2=k^2+l^2$ and {\bm $\sigma= \sigma_R + \rm i \sigma_I = \sigma_R + \rm i \omega$} is the complex growth rate. 
$S(z)$ is the amplitude function for $s$, $W(z)$ is the amplitude function for $w$ and $Z(z)$ is the amplitude function for the vertical vorticity $\zeta=\frac{\partial v}{\partial x} - \frac{\partial u}{\partial y}$. Taking the $z$-component of the curl and double-curl of the momentum equation to eliminate the pressure, along with the entropy equation, results in the following equations for the amplitude functions $W$, $Z$ and $S$:
\begin{align}\label{Ch4eq:GE2}
  \sigma Z=&Ta^{\frac{1}{2}}Pr\left[\sin\phi\left(DW+\frac{m\theta}{1+\theta
z}W\right)+\cos\phi \imi lW\right]\nonumber \\
&+Pr(D^2-a^2)Z+\frac{Prm\theta}{1+\theta
z}DZ,
\end{align}
\begin{align}
& -\sigma[D^2W-a^2W+\frac{m\theta}{1+\theta z}DW-\frac{m\theta^2}{(1+\theta
z)^2}W]=RaPra^2S \nonumber\\ 
 +&Ta^{\frac{1}{2}}Pr[\sin\phi DZ+\frac{m\theta}{1+\theta
z}\cos\phi \imi kW+\cos\phi \imi l
Z]-PrD^4W+2Pra^2D^2W \nonumber\\
 -&Pra^4W+\frac{3Prm(2-m)\theta^4}{(1+\theta z)^4}W
+\frac{2Prm\theta}{1+\theta z}a^2DW+\frac{2Prm^2\theta^2}{3(1+\theta
z)^2}a^2W \nonumber\\
 -&\frac{2Prm\theta}{1+\theta z}D^3W+\frac{Pr
m(4-m)\theta^2}{(1+\theta z)^2}D^2W-\frac{3Prm(2-m)\theta^3}{(1+\theta
z)^3}DW,\label{Ch4eq:GE1}
\end{align}
\begin{eqnarray}\label{Ch4eq:GE3}
  \sigma S-\frac{W}{1+\theta z}=\frac{1}{(1+\theta
z)^m}(D^2-a^2)S+\frac{\theta}{(1+\theta z)^{m+1}}DS,
\end{eqnarray}
to be solved subject to the boundary conditions:
\begin{equation}
 S=0, \text{\quad} W=0, \text{\quad} DZ=0 \text{\quad and \quad} D^2W+\frac{m\theta
DW}{(1+\theta z)}=0 \quad \text{on $z=0,1$}.
\end{equation}
Note, to simplify notation, we have used $D^n$ to mean $\frac{d^n}{dz^n}$. We solve this linear eigenvalue problem using the built-in bvp4c solver of Matlab.

Since we assume a reference state that is close to being adiabatic, we use $m=1.495$ in all calculations. \citet{Berkoffetal2010} demonstrated that the anelastic approximation gives a good approximation to fully compressible calculations even when the reference state is super-adiabatic, finding a $2\%$ error even when $\epsilon\sim10$; but as mentioned above, we prefer to remain close to the adiabatic state, where $m=1.5$.

%
\subsection{Effect of $\chi$ on the onset of convection}\label{linressub}
\begin{figure}[phtb!]
\begin{center}
\includegraphics[trim = 0mm 0mm 0mm 0mm, clip,scale=1]{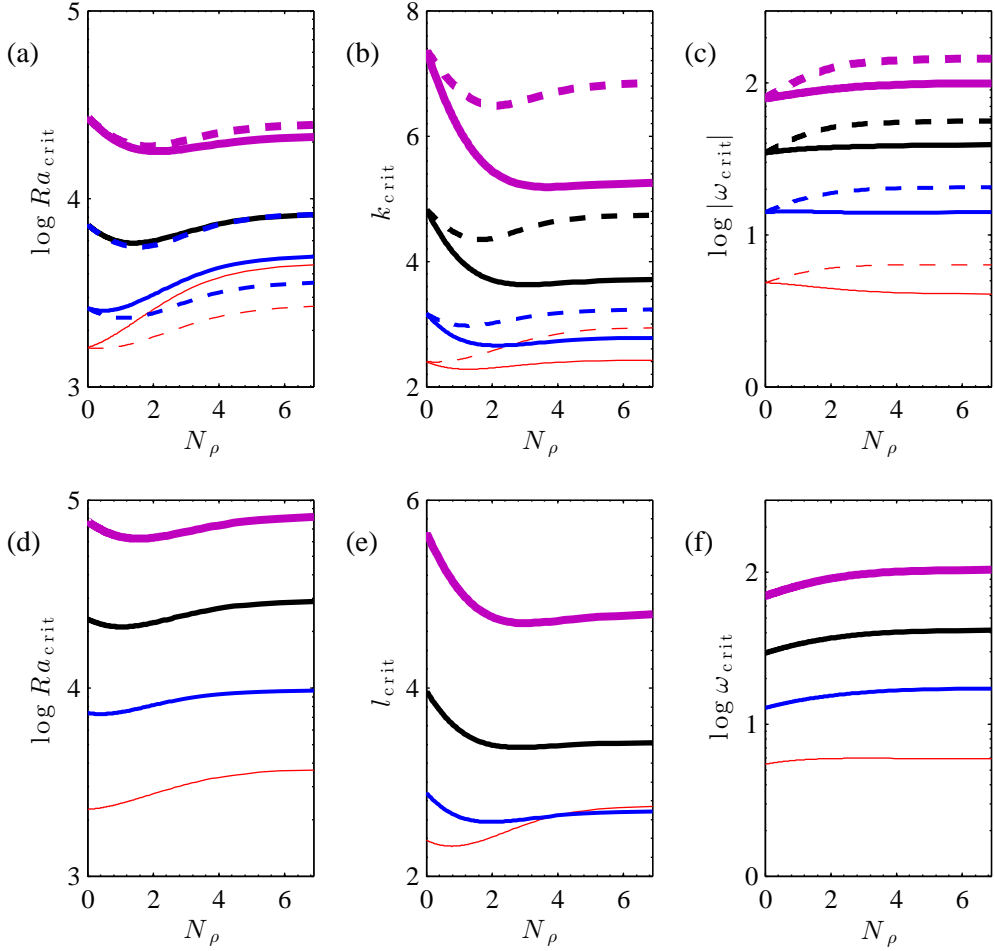}
\caption{Critical Rayleigh number (a), (d), wavenumber (b), (e) and frequency (c), (f) against $N_{\rho}$ for NS (a)-(c) and EW (d)-(f) rolls when $Pr=0.1$, \phiby4. Solid lines represent
solutions with $\omega_{\text{crit}}>0$ and dashed lines represent solutions with
$\omega_{\text{crit}}<0$. In red (thinnest line) $Ta=10^4$, in blue (thin line) $Ta=10^5$, in black (thick line) $Ta=10^6$ and in
purple (thickest line) $Ta=10^7$.}\label{linanelfig:RavthNS}
\end{center}
\end{figure}

Figure \ref{linanelfig:RavthNS} (a)-(c) show the critical Rayleigh number, wavenumber and frequency against $N_{\rho}$ for north-south (NS) convection rolls with $Pr=0.1$ and \phiby4. NS convection rolls are those whose axis is aligned in the $y$-direction ($l=0$) and similarly, east-west (EW) convection rolls are those whose axis is aligned in the $x$-direction ($k=0$). These are distinct when $\phi\neq\frac{\pi}{2}$, i.e., when rotation is oblique to gravity. For NS rolls, we observe that solutions with a positive frequency and those with a negative frequency are distinct unlike in the Boussinesq case \cite{Chandrasekhar} - we will examine this symmetry breaking in more detail in the next section. 
For small $Ta$, the negative branch is preferred but this changes to the positive branch as $Ta$ is increased. For the cases shown, the positive solution always has the smaller critical wavenumber and critical frequency. In addition, the minimum $Ra_{\text{crit}}$ occurs for $N_{\rho}>0$, with both the minimum $Ra_{\text{crit}}$ and the $N_{\rho}$ at which it occurs, increasing with $Ta$. Another feature of figure \ref{linanelfig:RavthNS} (a)-(c) is the difference between the positive and negative frequency solutions; this can be seen in the larger variation of $k_{\text{crit}}$ with increasing $N_{\rho}$ for the positive case.
In a similar study, \citet{Calkins2014} found that the critical Rayleigh number is a monotonic function of stratification. This difference could arise because of a number of different modelling choices. First, as discussed in the introduction, \citet{Calkins2014} take constant $\mu$ where we have constant $\nu$. Second, the parameter regime considered is very different to that covered here; they choose to focus on the rapidly rotating case ($Ta \sim 10^{10}$) at $Pr=1$. Finally, and perhaps most importantly, the definition of $Ra$ is different between our models (the difference resulting from the fact that we take $z$ pointing upwards and $\nu$ constant in contrast to \citet{Calkins2014} who take $z$ pointing downwards and $\mu$ constant.) Indeed, \citet{Calkins2014} do comment that their critical Rayleigh number is not a monotonic function of stratification if evaluated away from the bottom boundary.
We note that for vertical rotation we should recover the linear results of \citet{Mizerski} and indeed we do (though we do not present that case here). These differences highlight the important role of stratification in modifying the convection since the results are sensitive to whether $\mu$ or $\nu$ is held constant in the model as well as where $Ra$ is defined.

Figure \ref{linanelfig:RavthNS} (d)-(f) show the equivalent to figure \ref{linanelfig:RavthNS} (a)-(c) for EW rolls. Now, $k=0$, and there is no distinction to be made between the solutions with positive and negative frequency as they have the same critical values, hence we only plot the positive frequency solutions (we discuss this hidden symmetry in more detail in the next section). The behaviour is very similar to that in the NS case, but $Ra_{\text{crit}}$ is higher in the EW case, so that NS rolls are preferred, in agreement with \citet{Calkins2014} and also with linear Boussinesq systems, e.g., \citet{HTG}.

\subsection{Symmetry considerations}\label{proof}
\begin{figure}[htb]
\begin{center}
\includegraphics[trim=0mm 0mm 0mm 0mm clip,scale=1]{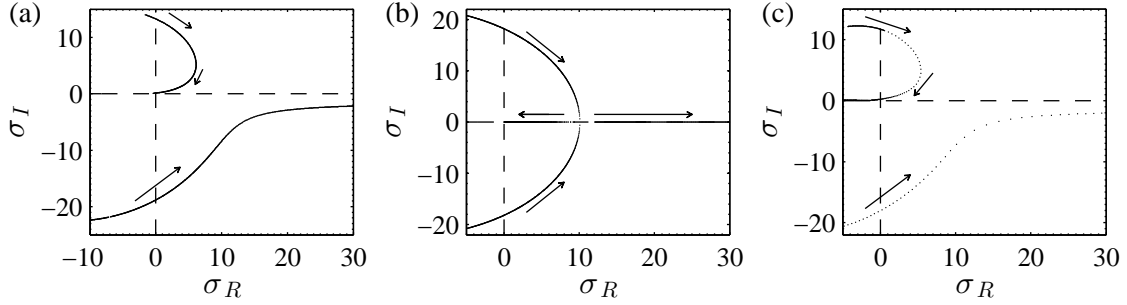}
\caption{Real and imaginary parts of growth rate plotted against each other for different $Ra$ whilst $Ta=10^5$, $Pr=0.1$, $\phi=\frac{\pi}{4}$, $\chi\approx 31$. In (a) $l=0$, $k=3$, in (b) $k=0$, $l=3$ and in (c) $k=3$, $l=3$. The arrows indicate the direction of increasing $Ra$. When $k=0$ there exists an unexpected symmetry in the eigenvalue spectrum.}\label{Ch4fig:bif123}
\end{center}
\end{figure}

As highlighted in the previous section, when $\chi\neq1$ ($N_{\rho}\neq 0$) and $l=0$ (NS rolls), there is a distinction to be made between solutions with a positive critical frequency and those with a negative critical frequency. However, when $k=0$ (EW rolls), even when $\chi\neq1$, there is still a symmetry and the positive and negative branches have the same $|\omega_{\text{crit}}|$. We might expect that breaking the up-down symmetry of the system, via the introduction of a vertical density stratification, would cause a break in symmetry of the eigenvalue spectrum, and hence result in different frequencies for the positive and negative branches. Instead, when $k=0$, the eigenvalues remain in complex conjugate pairs. We see that in figure \ref{Ch4fig:bif123} (a) and (c), the introduction of a vertical stratification across the layer has, as expected, broken the symmetry of the eigenvalue spectrum - they no longer appear in complex conjugate pairs. However, counter-intuitively, when $k=0$ (subfigure(b)), the symmetry is not broken and the eigenvalues remain in complex conjugate pairs, in an analogous way to the Boussinesq case ($\chi=1$). \citet{Evonuk2008} and \citet{Glatzmaieretal2009} describe a mechanism that is perhaps responsible for this difference between NS and EW rolls. The crux of their argument is that the vorticity equation (curl of equation (\ref{newmom})) contains a term proportional to $\bm\Omega(\nabla\cdot\mathbf u)$, which is in general, non-zero for anelastic convection. However, in our system, the $x$-component of this term is zero and so it does not have an effect on EW rolls, whereas, the $y$-component of this term is non-zero and so it does have an effect on NS rolls.

\begin{figure}[bht]
\begin{center}
\includegraphics[trim = 0mm 0mm 0mm 0mm clip,scale=1]{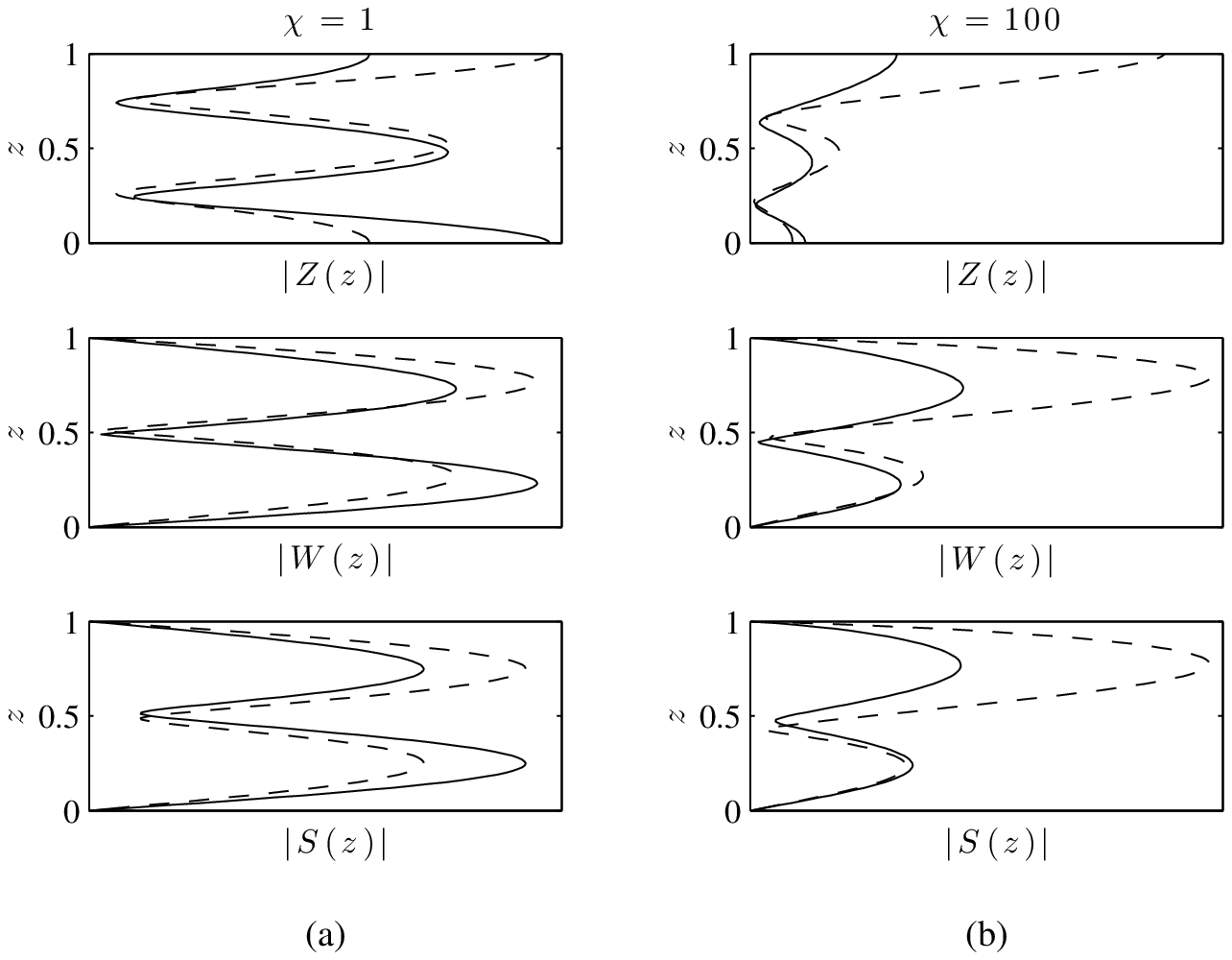}
\caption{Eigenfunctions. The solutions $|W(z)|$, $|Z(z)|$ and $|S(z)|$ as a function of $z$ for $k=0$, $l=2$, $Ta=10^5$, $Pr=0.1$, $\phi=\frac{\pi}{4}$, $Ra=2\times10^5$ and (a) $\chi=1$, (b) $\chi=100$. The solid line corresponds to the solutions with $\omega>0$ and the dotted lines to solutions with $\omega<0$.}\label{Ch4fig:eigf}
\end{center}
\end{figure}

To investigate the symmetry of the EW solutions further, we look at the eigenfunctions, $|W(z)|$, $|Z(z)|$ and $|S(z)|$ as a function of depth as in  figure \ref{Ch4fig:eigf}. The eigenvalues, as explained before, are a complex conjugate pair for both $\chi$; in (a) $\sigma=8.0489\pm 11.3672\imi$ and in (b) $\sigma=4.8626\pm 17.1070\imi$. It is clear from the plots that, in the Boussinesq case, (a), the eigenfunctions are symmetric about $z=0.5$, whereas when a stratification is added, (b), the corresponding eigenfunctions possess no obvious symmetry, despite the fact the eigenvalues are a complex conjugate pair.
This result is non-intuitive and so we give a proof of the maintenance of the symmetry when $k=0$. Essentially, the proof consists of forming the adjoint problem and then showing that the eigenvalue spectrum is symmetric; since the details are technical, they are included in the Appendix \ref{appendixproof}.

%
%
\section{Nonlinear results}\label{sec:NL}
This section extends the work of section \ref{sec:linear} to the nonlinear regime, allowing us to examine the mean flows driven by the system. For these nonlinear results we restrict ourselves to the 2d system which lies in the plane of the rotation vector, the $y$-$z$ plane, i.e., $\frac{\partial}{\partial x}\equiv 0$.

 We solve the nonlinear equations (\ref{newmom})-(\ref{news}) using a streamfunction, $\psi(y,z)$, defined by
\begin{equation}
\rho_{\rm{ref}}\bm u =\rho_{\rm{ref}} u \bm{\hat x}+\nabla \times \psi(y,z) \bm{\hat x}
=\left(\rho_{\rm{ref}} u, \frac{\partial \psi }{\partial z}, -\frac{\partial \psi}{\partial y}\right).
\end{equation}
and so $\nabla\cdot(\rho_{\rm{ref}}\mathbf{u})=0$ is automatically satisfied.
We then write the equations in terms of $\psi$ and $$\omega\equiv\nabla\times\mathbf u \cdot \hat x=-\frac{\nabla^2\psi}{\rho_{\rm{ref}}}-\frac{d}{dz}\left(\frac{1}{\rho_{\rm{ref}}}\right)\frac{\partial \psi}{\partial z},$$ and solve them for $\omega$, $u$, $s$ and $\psi$ using a Fourier-Chebyshev pseudospectral method with a second order, semi-implicit, Crank-Nicolson/Adams-Bashforth time-stepping scheme; for details see \citet{mythesis} and references within. $v$ and $w$ are then straightforward to obtain from $\psi$.

For diagnostic purposes, we decompose $\omega$, $u$, $s$ and $\psi$ into means (horizontal averages) and fluctuations; where the mean (denoted by an overbar) is defined as, for example,
 \begin{equation}
  \bar u(z,t)=\frac{1}{L}\int_0^Lu(y,z,t) \, dy, 
 \end{equation}
where $L$ is the length of the computational domain.

%
\subsection{Bifurcation structure and large-scale solutions}
\begin{figure}[ht]
\begin{center}
\includegraphics[trim = 0mm 0mm 0mm 0mm clip]{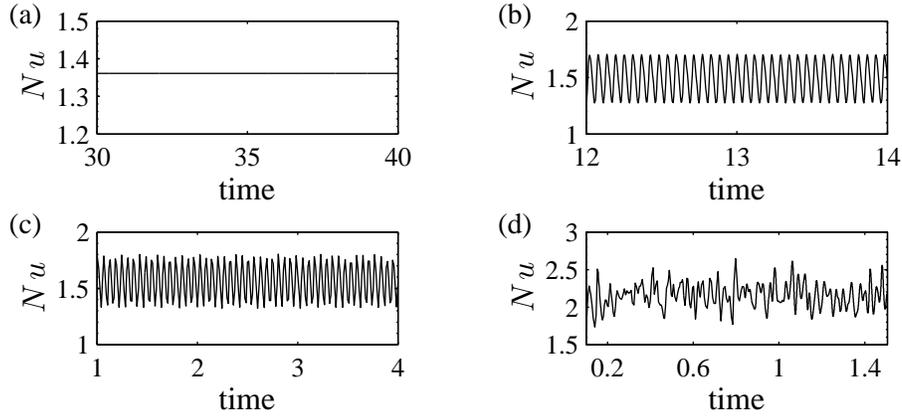}
\caption{Time series of Nusselt number ($Nu$) for the case when $Pr=1$, $Ta=10^5$, $\phi=\frac{\pi}{4}$, $\chi=5$ and (a) $Ra=4\times10^4$, (b) $Ra=4.2\times10^4$, (c) $Ra=4.6\times10^4$, (d) $Ra=7.5\times10^4$.}\label{fig1:chaos}
\end{center}
\end{figure}

In the limit $\theta \rightarrow 0$ ($\chi \rightarrow 1$), our anelastic system reduces to a Boussinesq system. As is typically seen in the Boussinesq case, if $Ra$ is slowly increased from its critical value (with other parameters fixed) then the solutions in the anelastic system undergo a series of bifurcations. An example is shown in figure \ref{fig1:chaos} which shows time series of the Nusselt number ($Nu$) at different $Ra$. We define $Nu$ to be the ratio of the total heat flux to the conductive heat flux in the basic state. Clearly the system undergoes a number of bifurcations via steady, oscillatory, quasiperiodic and chaotic solutions, en route to chaos (Ruelle-Takens-Newhouse route to chaos \cite{RuelleTakens1971, Newhouseetal1978}). We note that some hysteresis of the solutions was observed, depending on the initial conditions used; however, a full investigation of the bistability is not examined here.

\begin{figure}[thb!]
\begin{center}
\includegraphics[trim = 0mm 0mm 0mm 0mm clip]{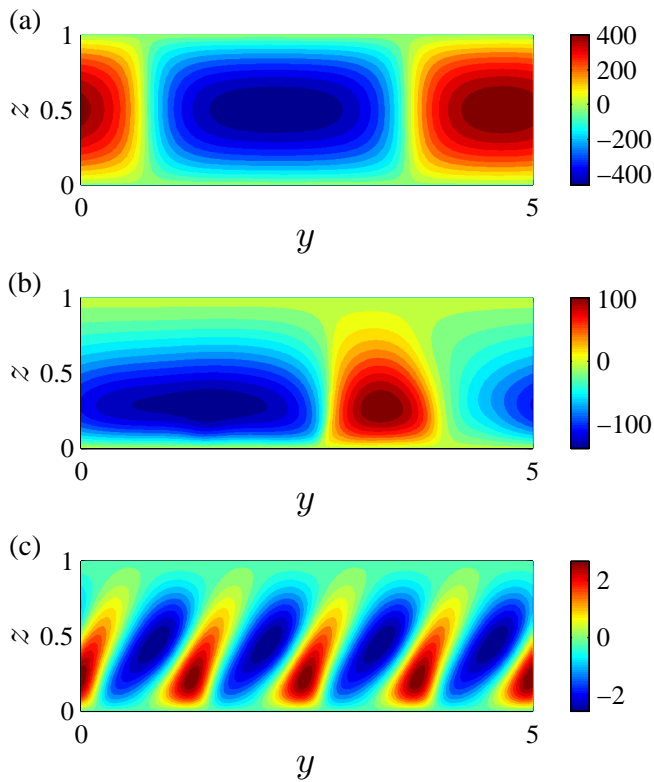}
\caption{ Contours of $\psi(y,z)$ for solutions when $L=5$, $Ra=8\times10^5$, $Ta=10^5$, $Pr=1$, $\phi=\frac{\pi}{6}$, (a) $\chi=1$ and (b) $\chi =10$. (c) shows $\psi(y,z)$ for the parameters in figure \ref{fig1:chaos} (a).}\label{fig:LSsoln}
\end{center}
\end{figure}

Whilst it is typical for the system to undergo this series of bifurcations, \citet{mythesis} reported on a regime in which steady, large-scale solutions that are efficient at transporting heat by convection are found to exist in Boussinesq convection. Large-scale here means the largest scale possible in a box of a given size. In the Boussinesq system such solutions have also been seen in non-rotating 2d Rayleigh-B\'enard convection\cite{ChiniCox2009, BenThesis2014}. 
In the anelastic system, we have found that such large-scale solutions are also able to exist even when the stratification is introduced; though interestingly, they may no longer correspond to steady (time-independent) solutions. For example, figure \ref{fig:LSsoln}(a) shows a snapshot of the large-scale steady solution that exists when $\chi=1$, i.e., in the Boussinesq limit and \ref{fig:LSsoln}(b) shows the equivalent large-scale solution when $\chi=10$. Whilst (a) corresponds to a steady solution, (b) is weakly time-dependent. For comparison, the dominant wavenumber of the equivalent solutions for the cases in figure \ref{fig1:chaos}, is about four times that of these large-scale solutions (e.g., see figure \ref{fig:LSsoln}(c)).
It is important to note that here we only consider $L=5$, but we would expect the width of the computational domain to have an effect on the emergence of the large-scale solutions. In general, a more detailed parameter study, which we do not carry out here, is required to examine more closely in which regimes such large-scale solutions exist.
Though the existence of large-scale solutions is of interest, the primary aim of this paper is to determine the role of stratification in modifying mean flows.

\subsection{Generation of mean flows}\label{subsec:anelmeanflows}
\begin{figure}[phtb!]
\begin{center}
\includegraphics{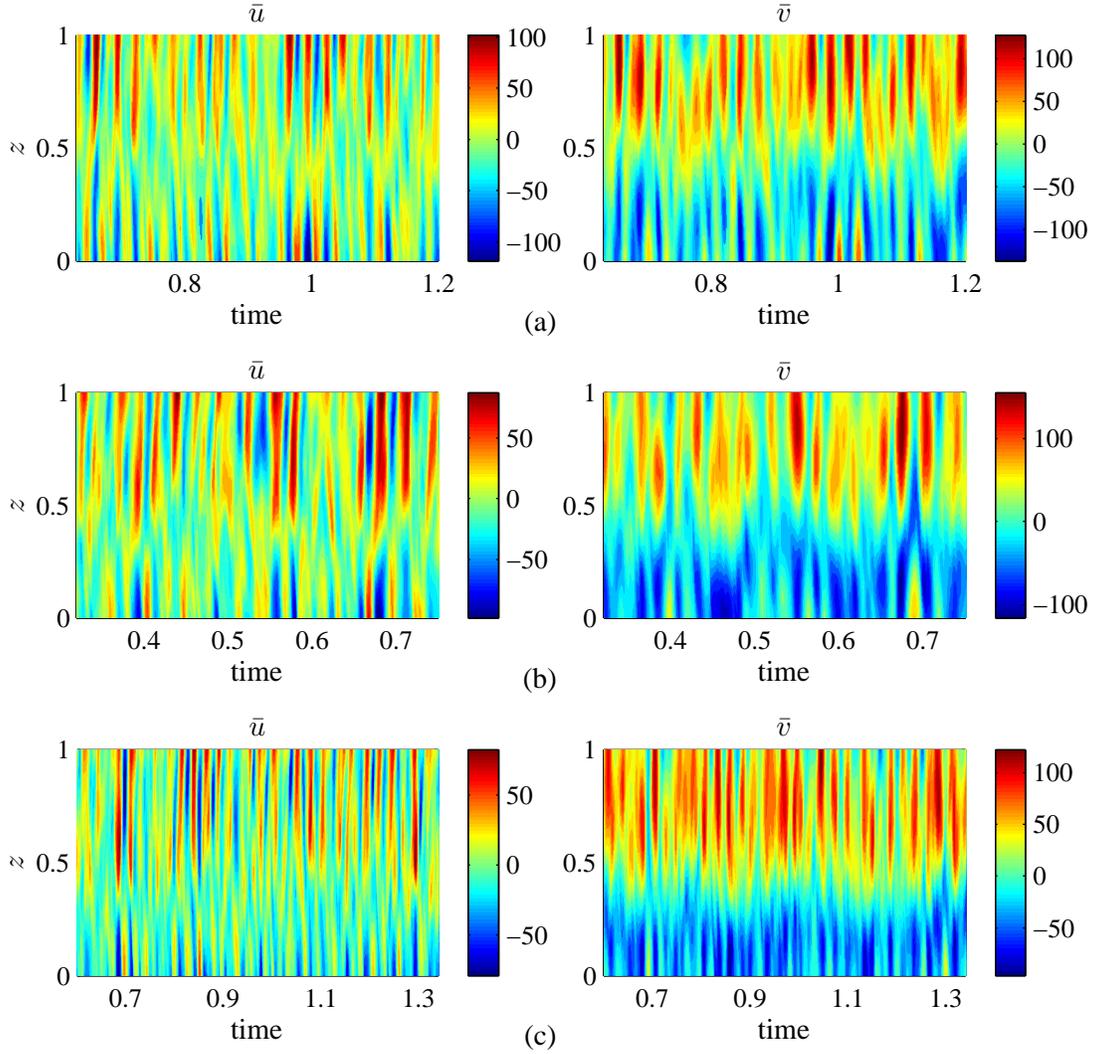}
\caption{Contour plots of the mean flows $\bar u$ and $\bar v$. In (a) $\chi=1.5$, in (b) $\chi=5$ and in (c) $\chi=10$. In all cases $Pr=1$, $Ra=2\times10^5$, $Ta=10^5$ and \phiby4.}\label{mfvtime}
\end{center}
\end{figure}

We investigate how the strength and direction of mean flows driven in our plane layer system are affected by stratification. In all simulations presented we fix the length of our computational box to be $L=5$.
To begin, we fix $Pr=1$, $Ra=2 \times 10^5$, $Ta=10^5$, \phiby4 and consider $\bar u$ and $\bar v$ for three different stratifications (see figure \ref{mfvtime}). In (a), $\chi=1.5$ (close to Boussinesq) in (b), $\chi=5$ and in (c), $\chi=10$. Even though the critical Rayleigh number changes with $\chi$ and $Ra$ is fixed in these cases, the degree of supercriticality is not vastly changed - ranging from approximately 5.8 to 6.2 times the onset value. 
A striking feature present in all of the flows shown in figure \ref{mfvtime} is that of strong oscillations. These oscillations are likely to be inertial oscillations that arise here because the Rossby number is of order one and so rotation and convection are of roughly equal importance\cite{Batchelor1967}. Such oscillations were also observed in the fully compressible calculations of \citet{Brummelletal1998}.
In figure \ref{mfvtime} (a), where $\chi$ is close to one, i.e., almost Boussinesq, we see that the flows are almost symmetric but, when the strength of stratification is increased, the extent of the asymmetry of the mean flows is also increased. For example, the positive flow of $\bar v$ in the upper half-plane only just penetrates down into the lower half-plane for $\chi$ close to one, but for stronger stratification it penetrates further into the layer. $\bar u$ is more time-dependent and harder to interpret than $\bar v$, but the asymmetry is still evident. The asymmetry results from the fact that, with stress free boundaries, the horizontal mass flux must be zero. 
From figure \ref{mfvtime}, other effects of increasing the stratification appear to be that the maximum velocity achieved by the flow decreases as $\chi$ increases, but the flows become more systematic. By systematic we mean a flow with a definite mean in the sense that the velocity fluctuations (including oscillations) produce a significant mean when averaged in both horizontal space and time.

Our results share some common features with those found in previous 3d studies. For example, we see strong systematic shear flows driven when the rotation vector is oblique to gravity as were seen in the Boussinesq calculations of \citet{HS1983} and the fully compressible calculations of \citet{Brummelletal1998}. Furthermore, the asymmetries introduced in the layer when stratification is added are also present in the flows of  \citet{Brummelletal1998}. We also find that $\bar u $ and $\bar v$ are comparable in size; a similar feature is also found in \citet{HS1983}  and \citet{Brummelletal1998}. This is likely to be a result of the horizontal periodic boundary conditions artificially enhancing the meridional flow. We also note that there are some differences between the flows driven here and those in the previous work of \citet{HS1983}  and  \citet{Brummelletal1998}. The most obvious difference being that the sense of $\bar u$ is reversed (whilst the sense of $\bar v$ coincides). We also tend to find that the percentage of energy in the mean flows compared to the total energy is larger in our cases. We expect these differences result from the difference in parameter regimes considered and also the 2d nature of our system. 

\begin{figure}[phtb!]
\begin{center}
 \includegraphics[scale=1]{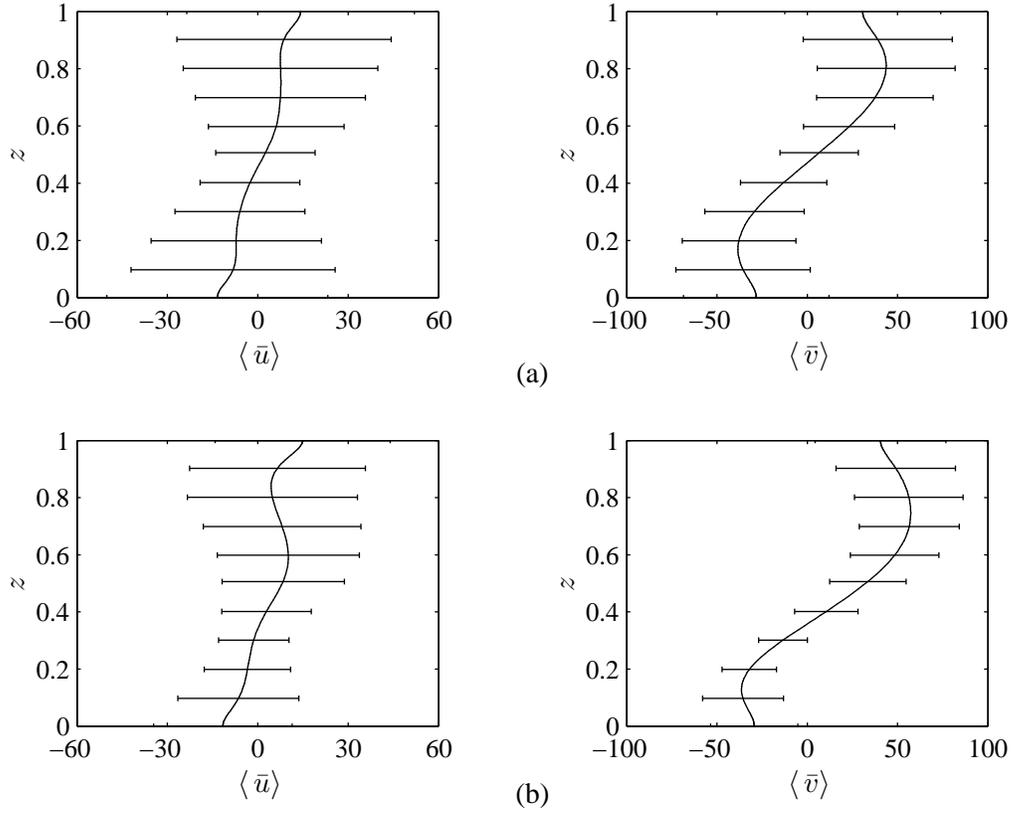}
\caption{Mean (black curve) and standard deviation (error bars) of $\langle\bar u\rangle$ and $\langle\bar v\rangle$ for $Pr=1$, $Ra=2\times10^5$, $Ta=10^5$, \phiby4 and (a) $\chi=1.5$, (b) $\chi=10$. As $\chi$ is increased the more systematic flow occurs at lower $z$.}\label{uvbar_herror}
\end{center}
\end{figure}

\begin{figure}[phtb!]
\begin{center}
\includegraphics{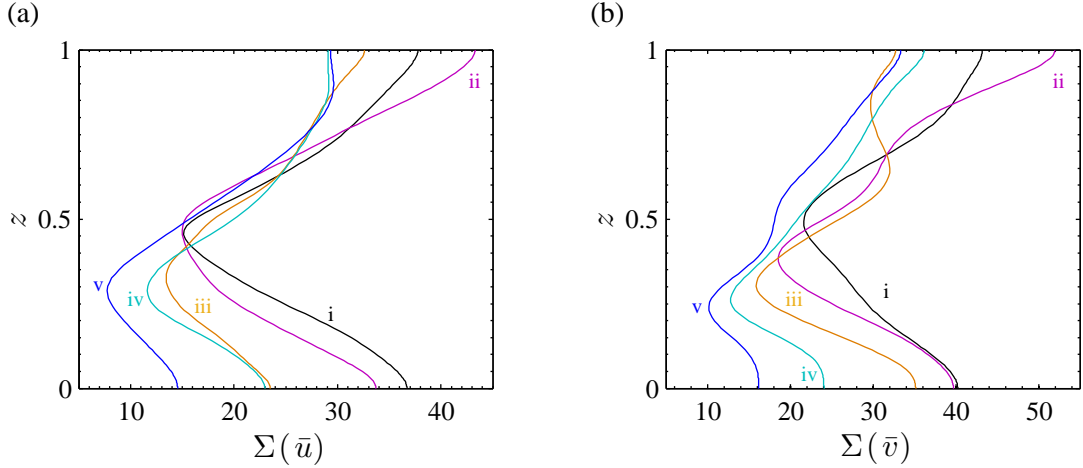}
\caption{Standard deviation of (a) $\bar u$ and (b) $\bar v$ as a function of layer depth for different stratifications. In black (i) $\chi=1.5$, in purple (ii) $\chi=2$, in orange (iii) $\chi=5$, in turquoise (iv) $\chi=10$ and in blue (v) $\chi=100$.}\label{sduvbar}
\end{center}
\end{figure}

To quantify the flow properties described above, we consider the mean and variation of the flows in time, and see how they vary with $\chi$ and $z$. In figure \ref{uvbar_herror}, we plot the time-averaged mean for $\bar u$ and $\bar v$ along with error bars corresponding to the standard deviation ($\Sigma$) from that mean. The first comment we make is that the error bars are significant and the departure of the maximum flow speed (given by the colorbars in figure \ref{mfvtime}) from the average is also significant; this is because of the oscillations present in the flows (as discussed above).
Further, in (a) we see that $\Sigma(\bar u)$ is smallest near to mid-layer and grows as we move out towards the boundaries but in (b), $\Sigma(\bar u)$ is smallest at a deeper layer. This behaviour is also seen in $\Sigma(\bar v)$, where for $\chi=1.5$, the standard deviation is fairly even across the layer but with its smallest value at approximately mid-layer; whereas for $\chi=10$, the smallest standard deviation is found at much smaller $z$. Note also, the mean of $\bar u$ and $\bar v$ is close to zero at $z=0.5$ in (a), but there is a significant flow at $z=0.5$ in (b). These measures characterise the behaviour we saw in the time-dependent plots in figure \ref{mfvtime}. As a percentage of its mean, $\Sigma(\bar u)$ is larger than $\Sigma(\bar v)$, indicative of the more time-dependent behaviour of $\bar u$ we also observed in figure \ref{mfvtime} .
Figure \ref{sduvbar} shows how the standard deviation varies with $z$ for a number of different $\chi$. In general, we see that for stronger stratification $\Sigma$ is reduced.  This behaviour is particularly evident in the lower depths of the layer (smaller z). 
 It is also evident that for $\Sigma(\bar v)$, the minimum of the standard deviation occurs at a deeper level in the layer as $\chi$ is increased. For $\Sigma(\bar u)$, the trend is not so clear, however, the flows corresponding to larger $\chi$ have a minimum at a lower $z$ than the flows corresponding to smaller $\chi$. Therefore, there are fewer fluctuations at lower levels with increasing $\chi$, and it is this that results in the relatively large time-averaged mean at this level.

Reynolds stresses are known to drive mean flows \cite{HS1983,Brummelletal1998}.  To analyse their role in mean flow generation, we consider the mean equations obtained by horizontally averaging the $x$ and $y$ components of the momentum equation, i.e.,
\begin{equation}\label{mfeq1a}
 Pr\rho_{\rm{ref}}\langle\bar u\rangle = \frac{Pr}{Ta^{\frac{1}{2}}\sin\phi}\frac{\partial}{\partial z}
\left(\rho_{\rm{ref}} \frac{\partial\langle\bar v\rangle}{\partial
z}\right)-\frac{1}{Ta^{\frac{1}{2}}\sin\phi}\frac{\partial
(\rho_{\rm{ref}}\langle\overline{vw}\rangle)}{\partial z} ,
\end{equation}
\begin{equation}\label{mfeq2a}
 Pr\rho_{\rm{ref}}\langle\bar v\rangle = -\frac{Pr}{Ta^{\frac{1}{2}}\sin\phi}\frac{\partial}{\partial z}
\left(\rho_{\rm{ref}} \frac{\partial\langle\bar u\rangle}{\partial
z}\right)+\frac{1}{Ta^{\frac{1}{2}}\sin\phi}\frac{\partial
(\rho_{\rm{ref}}\langle\overline{uw}\rangle)}{\partial z}, 
\end{equation}
where we have averaged in time and assumed a statistically steady state so that $\frac{\partial }{\partial t}\langle \bar u \rangle=\frac{\partial }{\partial t}\langle \bar v \rangle=0$. 
The quantities $\rho_{\rm{ref}}\overline{uw}$, $\rho_{\rm{ref}}\overline{vw}$ are the Reynolds stresses terms, they measure the correlation between the horizontal and vertical velocity components. With a tilted rotation vector we might expect these correlations to be nonzero\cite{HS1983}. We note, from equations (\ref{mfeq1a}) and (\ref{mfeq2a}), that it is the $z$-derivative of $\rho_{\rm{ref}}\overline{vw}$ that drives $\bar u$ and the $z$-derivative of $\rho_{\rm{ref}}\overline{uw}$ that drives $\bar v$. In what follows, for both equations (\ref{mfeq1a}) and (\ref{mfeq2a}), we refer to the term on the LHS as the Coriolis term, the first term on the RHS as the viscous term and the second term on the RHS as the Reynolds stress (RS) term.

\begin{figure}[phtb!]
\begin{center}
\includegraphics{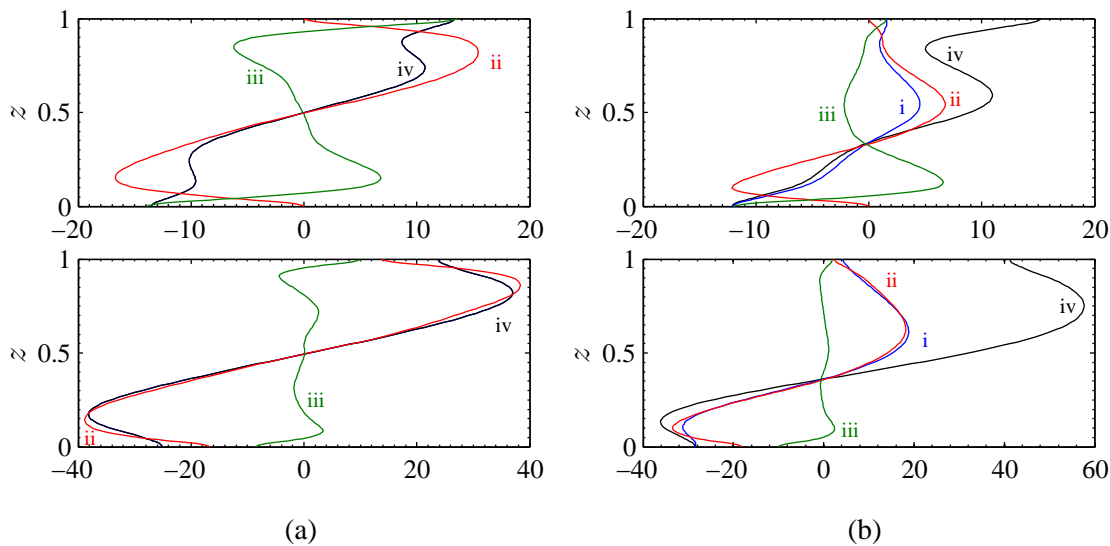}
\caption{Terms of the mean flow equations (\ref{mfeq1a}) (top panels) and (\ref{mfeq2a}) (bottom panels) as a function of $z$ for $Pr=1$, $Ra=2\times10^5$ and in (a), $\chi=1$, whilst in (b), $\chi=10$. The blue lines (i) represent the Coriolis terms, the red (ii) the RS terms, the green (iii) the viscous terms and black (iv) the mean flows $\langle\bar u\rangle$ and $\langle\bar v\rangle$. In case (a), the Coriolis terms are equivalent to the mean flows themselves.}\label{termsvth}
\end{center}
\end{figure}

The factor of $\rho_{\rm{ref}}$ in the Coriolis terms of equations (\ref{mfeq1a}) and (\ref{mfeq2a}) means that, in theory, for two different $\chi$, if the driving terms on the right-hand side are of the same size, then the case with the largest $\chi$ will yield the largest $\bar u$ and $\bar v$, i.e., at any fixed $z$, if $Pr \rho_{\rm{ref}} \bar u$ is the same for two different $\rho_{\rm{ref}}$ (fixed $Pr$) then $\bar u$ will be larger for the smaller $\rho_{\rm{ref}}$. To see this, we plot each of the terms of equations (\ref{mfeq1a}) and (\ref{mfeq2a}) in figure \ref{termsvth}; in addition, we plot $\bar u$ and $\bar v$ (without the $Pr\rho_{\rm{ref}}$ factors). For $\chi=1$, the Coriolis term is equivalent to the mean flow since $Pr=1$, therefore no additional line is visible in this case. However, for $\chi\neq1$, there is a difference between the Coriolis term and the mean flow itself. In both (a) and (b), the strong dominance of the RS terms is clear. It is also evident that the viscous term is more important in determining $\bar u$ than it is $\bar v$, this is because the viscous term affecting each mean flow component depends on the gradient of the other mean flow component and we find that the gradient of $\bar v$ tends to be larger than that of $\bar u$.  It is clear that the RS terms are larger in the $\chi=1$ case and this results in the Coriolis terms being larger for $\chi=1$. However, because for $\chi=10$, $\rho_{\rm{ref}}\leq 1$ across the layer, $\langle\bar u\rangle$ and $\langle\bar v\rangle$ are actually larger for $\chi=10$. This effect is most prominent at the top of the layer, where the fluid mass is at its lowest.

In the Boussinesq case, the RS terms are symmetric about the mid-layer depth. As a result of this symmetry, and because the RS terms are the dominant terms in equations (\ref{mfeq1a}) and (\ref{mfeq2a}), the mean flows are also symmetric about the mid-layer depth. Moreover, if $\chi$ is increased, then the RS terms become asymmetric, leading to asymmetric mean flows. We comment that the symmetry of the RS terms can change both as a result of the small-scale turbulent interactions but also through the mean flow acting back on the turbulence.

\section{Conclusion}\label{sec:Conclusion}
The results in this paper can be categorised into two distinct sections. The first section was concerned with linear anelastic convection. There we demonstrated and proved the existence of a previously unknown, hidden symmetry in the equations present when EW rolls are considered. We explained this by showing that the symmetry breaking term present in the vorticity evolution equation (introduced by the inclusion of a stratification) vanishes under certain circumstances.

The second part of the paper was concerned with the effect of stratification on nonlinear anelastic convection. We showed that efficient large-scale convection cells that have been shown to exist in 2d, Boussinesq Rayleigh-B\'enard convection can also be found when stratification is introduced; although, unlike their Boussinesq counterparts, the anelastic solutions may not be time independent.

We went on to examine the effect of stratification on the generation of zonal and meridional mean flows. The use of an idealised 2d tilted plane layer model has allowed us to show the importance of correlations between velocity components resulting in Reynolds stresses that generate systematic flows. These flows have a strong time dependence but have a definite preferred direction on time averaging.
The most striking difference between the flows driven in a stratified layer and their Boussinesq counterparts is seen in a vertical asymmetry. Although flow velocities tended to decrease with $\chi$, a statistical analysis showed that the mean flows become more systematic at lower layer depths the stronger the stratification.
The asymmetry introduced by stratification is seen in the Reynolds stresses and we highlighted the role these stresses play in determining the size and vertical structure of the mean flows. In particular, even though the RS terms can be larger in the Boussinesq case, the flows are faster in the anelastic case because of the reduced fluid density in the layer. 
Whilst the Reynolds stresses dominated the flow size and structure, the viscous forces played a role in modifying the mean flows by opposing the Reynolds stresses (an effect that is to be expected at $Pr=1$). However, we would expect the viscous forces to be much less important in a realistic setting as the diffusivities are much smaller.
In fact, \citet{mythesis} found small $Pr$ to play an important part in the dynamics of mean flow generation in Boussinesq convection. In particular, for the same size Reynolds stresses a larger mean flow is driven at smaller $Pr$. We would expect a similar effect to occur here as, from equations (\ref{mfeq1a}) and (\ref{mfeq2a}), the RS terms drive $Pr \rho_{\rm ref} \langle \bar u \rangle$ and $Pr \rho_{\rm ref}\langle \bar v \rangle$ and so for smaller $Pr$,  $\rho_{\rm ref} \langle \bar u \rangle$ and  $\rho_{\rm ref} \langle \bar v \rangle$ are indeed larger at any fixed height. The role of small $Pr$ in anelastic convection is currently under investigation.

There are some obvious shortcomings resulting from the simplicity of our model. For example, the periodic boundary conditions unrealistically enhance the meridional flow when in actuality, zonal flows are usually much larger than the meridional ones, in e.g., the Sun or on Jupiter.
Despite this, our simple analysis has shown the importance of the Reynolds stresses in mean flow generation and shed some light on the role of stratification.

Owing to the existence of magnetic fields in physical systems such as stars and planets and their interaction with convection and rotation, an obvious question to ask is how such magnetic fields modify the mean flows generated. This is a question we address in a subsequent paper.
We conclude by acknowledging the limitations of our simple model. Clearly in two dimensions correlations may be over-exaggerated leading to the formation of strong mean flows. We are therefore currently investigating how extending the model to three dimensions weakens correlations and affects the turbulent driving of mean flows.

%
%
%
%
%
%
%
%
%
%
%
%
%
%
%
%
%
%
%
%
%
%
%
%
%
%
%
%
%
%
%
%
%
%
%
%
%
%
%
%
%
%
%
%
%
\begin{acknowledgments}
L.K.C. is grateful to Science and Technology Facilities Council (STFC) for a PhD studentship. We thank the anonymous referees for comments that helped to focus the content of this paper.
\end{acknowledgments}
%

\appendix*
\section{Proof}\label{appendixproof}
The following is a proof of the symmetry of the spectrum of eigenvalues that exists when
$k=0$ (see section \ref{proof}). The proof not only holds for the stress free boundary conditions considered above
but is a more general result and holds for all natural boundary conditions. The proof is
similar in nature to that of \citet{Proctor2011} who prove a similar result. However, they
consider a system with symmetric equations but break the symmetry through asymmetric
boundary conditions. This is in contrast to this work, where we have asymmetric equations
to begin with, and typically our boundary conditions are symmetric.

To begin the proof, we make a change of variables. Let
\begin{align}
 \tilde Z&=(1+\theta z)^{\frac{m}{2}}Z,\label{cofv1}\\
 \tilde W&=(1+\theta z)^{\frac{m}{2}}W, \\
 \tilde S&=(1+\theta z)^{\frac{1}{2}}S,\label{cofv3}
\end{align}
then multiply (\ref{Ch4eq:GE2}) and (\ref{Ch4eq:GE1}) by $(1+\theta z)^{\frac{m}{2}}$,
 (\ref{Ch4eq:GE3}) by $({1+\theta z})^{m+\frac{1}{2}}RaPra^2$ and substitute in
(\ref{cofv1}) - (\ref{cofv3}), to give
\begin{align}
  \sigma\tilde Z=&Ta^{\frac{1}{2}}Pr\left[\sin\phi\left(D\tilde
W+\frac{m\theta}{2(1+\theta
z)}\tilde W\right)+\cos\phi \imi l \tilde W\right]+\nonumber\\&Pr(D^2-a^2)\tilde
Z-\frac{Pr m\theta^2(\frac{m}{2}-1)}{2(1+\theta z)^2}\tilde Z,
\end{align}
\begin{align}\label{Ch4eq:red}
& -\sigma[(D^2-a^2)\tilde W-\frac{m\theta^2(1+\frac{m}{2})}{2(1+\theta
z)^2}\tilde W]=RaPra^2(1+\theta z)^{\frac{m-1}{2}}\tilde S\nonumber\\
&+Ta^{\frac{1}{2}}Pr\sin\phi[D\tilde Z-\frac{m\theta}{2(1+\theta
z)}\tilde Z]+Ta^{\frac{1}{2}}Pr\cos\phi \rm{i} l\tilde Z \nonumber\\
&-PrD^4\tilde W+2Pra^2D^2\tilde W-Pra^4\tilde W+ \frac{Pr
m\theta^2(\frac{m}{2}+1)}{(1+\theta z)^2}D^2\tilde W \nonumber\\
&-\frac{Pr\theta^3m(m+2)}{(1+\theta z)^3}D\tilde W +\mathcal{F\tilde W} + Ta^{\frac{1}{2}}Pr\cos\phi\frac{m\theta}{1+\theta z}\imi k\tilde W,
\end{align}
where
\begin{align}
 \mathcal{F}&=\frac{Prm\theta^4(3+\frac{5m}{4}-\frac{m^2}{4}-\frac{m^3}{16})}{(1+\theta
z)^4}+\frac{Prma^2\theta^2(1+\frac{m}{6})}{(1+\theta
z)^2},
\end{align}
and
\begin{align}
   \sigma RaPra^2(1+\theta z)^m\tilde S=RaPra^2(1+\theta
z)^{\frac{m-1}{2}}\tilde W+RaPra^2(D^2-a^2)\tilde S+\frac{RaPra^2\theta^2}{4(1+\theta
z)^2}\tilde S.
\end{align}
When $k=0$, $a=l$ and we can write this system as
\begin{equation}
 \sigma \mathbf{A}\mathbf{\tilde{X}}=\mathbf{B}\mathbf{\tilde{X}}
\end{equation}

where $\mathbf{\tilde{X}}=\begin{bmatrix}
       \tilde Z\\
       \tilde W\\
       \tilde S  
      \end{bmatrix},$
$\mathbf{A}=\begin{bmatrix}
       1 & 0 & 0\\
       0 & -(D^2-l^2)+\frac{m\theta^2(\frac{m}{2}+1)}{2(1+\theta z)^2} & 0\\
       0 & 0 & RaPrl^2(1+\theta z)^m
      \end{bmatrix}$ and $\mathbf{B}=$

\begin{footnotesize}
 $$\begin{bmatrix}
\multirow{2}{*}{$Pr(D^2-l^2)-\frac{Pr m \theta^2(\frac{m}{2}-1)}{2(1+\theta z)^2}$} &
Ta^{\frac{1}{2}}Pr[\sin\phi(D+\frac{m\theta}{2(1+\theta z)}) & \multirow{2}{*}{$0$}\\
& +\cos \phi \imi l] &\\
& & \\
 Ta^{\frac{1}{2}}Pr[\sin\phi(D-\frac{m\theta}{2(1+\theta z)}) &
 -Pr(D^2-l^2)^2+\frac{Prm\theta^2(\frac{m}{2}+1)}{2(1+\theta z)^2}D^2
&\multirow{2}{*}{$RaPrl^2(1+\theta z)^{\frac{m-1}{2}}$}\\
+\cos \phi \imi l]&-\frac{Pr\theta^3m(m+2)}{(1+\theta z)^3}D+\mathcal{F}& \\
& & \\
0 & RaPrl^2(1+\theta z)^{\frac{m-1}{2}} & RaPrl^2[(D^2-l^2)+\frac{\theta^2}{4(1+\theta
z)^2}]
      \end{bmatrix}.$$
\end{footnotesize}

Next, we define the inner product
\begin{equation}
 \langle \mathbf{\tilde X_1},\mathbf{\tilde X_2} \rangle=\int_{0}^{1} \mathbf{\tilde
X_1}^{{*}^{T}}\mathbf{\tilde X_2}
\, dz = \int_{0}^{1} (\mathbf{\tilde X_2}^{{*}^{T}}\mathbf{\tilde X_1})^*\, dz=\langle
\mathbf{\tilde X_2},\mathbf{\tilde X_1} \rangle^*
\end{equation}
where \begin{equation}
     \mathbf{\tilde X_1}=\begin{bmatrix}
       \tilde Z_1\\
       \tilde W_1\\
       \tilde S_1  
      \end{bmatrix}, \hspace{5mm}
     \mathbf{\tilde X_2}=\begin{bmatrix}
       \tilde Z_2\\
       \tilde W_2\\
       \tilde S_2  
      \end{bmatrix},
    \end{equation}
and $\mathbf{\tilde X_1}$ satisfies the same boundary conditions as $\mathbf{\tilde X_2}$.
Then, since $\mathbf{A}$ is real and symmetric,
\begin{align}\label{Ch4eq:IP1}
 \langle \mathbf{\tilde X_1},(\sigma \mathbf{A}\mathbf{\tilde
X_2}-\mathbf{B}\mathbf{\tilde X_2}) \rangle&=\int_{0}^{1}
\mathbf{\tilde X_1}^{{*}^{T}}(\sigma \mathbf{A}\mathbf{\tilde
X_2}-\mathbf{B}\mathbf{\tilde X_2})\, dz \nonumber\\
& =\int_{0}^{1} \mathbf{\tilde X_2}^{T}(\sigma^{*} \mathbf{A}\mathbf{\tilde
X_1}-\mathbf{B}^\dagger\mathbf{\tilde X_1})^*\, dz
= \langle (\sigma^* \mathbf{A}\mathbf{\tilde X_1}-\mathbf{B}^\dagger\mathbf{\tilde
X_1}),\mathbf{\tilde X_2} \rangle.
\end{align}
Note, equation (\ref{Ch4eq:IP1}) only holds if the boundary conditions on $\mathbf{\tilde
X_i}$ and $\mathbf{\tilde X_i}^*$ $(i=1,2)$ are the same. 
So $\mathbf{B}^\dagger$ is the formal adjoint of
$\mathbf{B}$, i.e., $\langle \mathbf{u},\mathbf{B}\mathbf{v} \rangle=\langle
\mathbf{B}^\dagger \mathbf{u}, \mathbf{v} \rangle$ for vectors $\mathbf{u}$ and
$\mathbf{v}$ and it is given by $\mathbf{B}^\dagger=$

\begin{footnotesize}
$\begin{bmatrix}
\multirow{2}{*}{$Pr(D^2-l^2)-\frac{Pr m \theta^2(\frac{m}{2}-1)}{2(1+\theta z)^2}$} &
-Ta^{\frac{1}{2}}Pr[\sin\phi(D+\frac{m\theta}{2(1+\theta z)}) & \multirow{2}{*}{$0$}\\
& +\cos \phi \imi l] &\\
& & \\
-Ta^{\frac{1}{2}}Pr[\sin\phi(D-\frac{m\theta}{2(1+\theta z)}) &
-Pr(D^2-l^2)^2+\frac{Prm\theta^2(\frac{m}{2}+1)}{2(1+\theta z)^2}D^2
&\multirow{2}{*}{$RaPrl^2(1+\theta z)^{\frac{m-1}{2}}$}\\
+\cos \phi \imi l]&-\frac{Pr\theta^3m(m+2)}{(1+\theta z)^3}D+\mathcal{F}& \\
& & \\
0 & RaPrl^2(1+\theta z)^{\frac{m-1}{2}} & RaPrl^2[(D^2-l^2)+\frac{\theta^2}{4(1+\theta
z)^2}]
\end{bmatrix}.$
\end{footnotesize}

Since $\mathbf{B}^\dagger$ is the formal adjoint of $\mathbf{B}$, its spectrum is the
complex conjugate of
the spectrum of B. Now, if we let 
\begin{equation}
 \mathbf{\tilde Y_1}=\begin{bmatrix}
       -\tilde Z_1\\
       \tilde W_1\\
       \tilde S_1  
      \end{bmatrix},
\end{equation}
then the adjoint equation \begin{align}\label{last1}
      \sigma^*\mathbf{A}\mathbf{\tilde X_1}&=\mathbf{B}^\dagger\mathbf{\tilde X_1} \quad
\text{can be written as }\\\label{last2}       \sigma^*\mathbf{A}\mathbf{\tilde
Y_1}&=\mathbf{B}\mathbf{\tilde Y_1} \quad \text{ when } k=0.
     \end{align}
So, if ($\sigma$, $\mathbf{\tilde X_1}$) is an eigenvalue, eigenfunction pair for the
system then so is ($\sigma^*$, $\mathbf{\tilde{Y}_1}$).

Hence, we have shown that as long as the boundary conditions on $\mathbf{\tilde{X}}$ and
$\mathbf{\tilde{X}}^*$ are the same, when $k=0$, the eigenvalue spectrum is
symmetric. This is in agreement with the numerical results we found in section \ref{linressub}.

If $k\neq0$, then the final term on the right-hand-side of equation
(\ref{Ch4eq:red}) is non-zero and must be added to the central entry of the matrices $\mathbf{B}$ and
$\mathbf{B}^\dagger$; this results in a breakdown of the proof, as the last step (from
equation (\ref{last1}) to equation (\ref{last2})) can not be carried out.
Therefore, when $k\neq0$, the eigenvalue spectrum is not symmetric, again in agreement
with the numerical results obtained in section \ref{linressub}.

\bibliographystyle{unsrtnat}
\bibliography{hydropaper3}
\end{document}